%% file: main.tex
\documentclass[conference]{IEEEtran}
\usepackage{amsfonts}
\usepackage{amssymb}
\usepackage{cite}
\usepackage[cmex10]{amsmath}
\usepackage{float}
\usepackage{color}
\usepackage{stfloats,fancyhdr}
\usepackage{amsmath}
\usepackage{amsthm}
\usepackage{algorithm}
\usepackage{algorithmic}
\usepackage{multirow}
\usepackage{changepage}
\usepackage[normalem]{ulem}

\usepackage{url}

\IEEEoverridecommandlockouts

\ifCLASSINFOpdf
  \usepackage[pdftex]{graphicx}
  \DeclareGraphicsExtensions{.pdf,.jpeg,.png}
\else
  \usepackage[dvips]{graphicx}
  \DeclareGraphicsExtensions{.pdf}
\fi

\usepackage{subfigure}

\usepackage{fancybox,dashbox}
\begin{document}
\title{{Software-Defined Radio Implementation of Age-of-Information-Oriented Random Access}
\thanks{$^\dag$The work of H. Chen is supported by the CUHK direct grant under the project code 4055126. Any technical issue regarding this paper should go to H. Chen.}
\thanks{$^*$The first two authors made equal contributions to this paper.}
%\thanks{H. Chen, J. Liang, Z. Han, and S. C. Liew are with Department of Information Engineering, The Chinese University of Hong Kong, Hong Kong SAR, China (email: \{he.chen, lj015, hz019, soung\}@ie.cuhk.edu.hk).}%zl018
%\thanks{Y. Gu, R. Wang, and Y. Li are with School of Electrical and Information Engineering, The University of Sydney, Sydney, NSW 2006, Australia (email: yifan.gu@sydney.edu.au, rwan7881@uni.sydney.edu.au, yonghui.li@sydney.edu.au).}
}
%\author{Zixiao Han$^1$, Jiaxin Liang$^1$, Yifan Gu$^2$, and He Chen$^1$}
\author{\IEEEauthorblockN{Zixiao Han\textsuperscript{1}$^{,*}$, Jiaxin Liang\textsuperscript{1}$^{,*}$, Yifan Gu\textsuperscript{2}, and He Chen\textsuperscript{1}$^{,\dag}$}%
\IEEEauthorblockA{\textsuperscript{1}Department of Information Engineering, The Chinese University of Hong Kong, Hong Kong SAR, China}
\IEEEauthorblockA{\textsuperscript{2}School of Electrical and Information Engineering, The University of Sydney, Sydney, Australia\\
\textsuperscript{1}\{hz019, lj015, he.chen\}@ie.cuhk.edu.hk, \textsuperscript{2}yifan.gu@sydney.edu.au}}

\maketitle
\vspace{-1em}
\begin{abstract}
More and more emerging Internet of Things (IoT) applications involve status updates, where various IoT devices monitor certain physical processes and report their latest statuses to the relevant information fusion nodes. %Due to the large number of devices in massive IoT networks, the decentralized channel access protocols (e.g. random access) are preferable. %Existing random access protocols have been largely designed and optimized for maximizing the network throughput or minimizing the packet delay, which have been shown to be suboptimal in status update systems emphasizing on information freshness and timeliness.
{A new performance measure, termed age of information (AoI), has recently been proposed to quantify the information freshness in time-critical IoT applications. Due to the large number of devices in future IoT networks, the decentralized channel access protocols (e.g. random access) are preferable thanks to their low network overhead.} Built on the AoI concept, {some recent efforts have developed several AoI-oriented ALOHA-like random access protocols for boosting the network-wide information freshness.} However, all relevant works focused on theoretical designs and analysis. The development and implementation of a working prototype to evaluate and further improve these random access protocols in practice  have been largely overlooked. Motivated as such, we build a software-defined radio (SDR) prototype for testing and comparing the performance of recently proposed AoI-oriented random access protocols. To this end, we implement a time slotted wireless system by devising a simple yet effective over-the-air time synchronization scheme, in which beacons that serve as reference timing packets are broadcast by an access point (AP) from time to time. For a complete working prototype, we also design the frame structures of various packets exchanged within the system. Finally, we design a set of experiments, implement them on our prototype and test the considered algorithms in an office environment.

%We then migrate our designs to another prototype built on off-the-shelf WiFi network cards to demonstrate the feasibility of the state-of-the-art AoI-oriented random access policies on commercial devices.

%This paper concentrates on the design, analysis, and implementation of a new age-oriented random access protocol for massive IoT networks, where existing centralized scheduling mechanisms are no longer applicable due to the large number of devices.
%Specifically, we devise a stationary threshold-based age-dependent random access (ADRA) protocol, in which each IoT device accesses the common channel with a certain probability only when its instantaneous age exceeds a predetermined threshold. We evaluate the average AoI of the proposed ADRA protocol mathematically by decoupling the tangled AoI evolutions of multiple IoT devices and modelling the decoupled AoI evolution of each device as a Discrete-Time Markov Chain. Simulation results validate our theoretical analysis and affirm the superior age performance of the proposed ADRA protocol over its conventional age-independent counterpart. To evaluate the proposed ADRA protocol in real scenarios, we build a proof-of-concept prototype using the USRP (universal software radio peripheral) platforms. Our experiment results corroborate our theoretical analysis.
\end{abstract}

%\begin{IEEEkeywords}
%Industrial Internet of Things, information freshness, status updates, age-of-information, random access, software-defined radio, implementation.
%\end{IEEEkeywords}


\IEEEpeerreviewmaketitle

\input{introduction}
\input{system_model}
\input{expriment}

% or

{
\section{Conclusions}
In this paper, we developed a time slotted wireless prototype by using the Ettus USRP software-defined radio platform for experimental evaluation and comparison of recently proposed age-of-information-oriented ALOHA-like random access protocols. Our experimental results showed that the age-dependent random access policy outperforms its age-independent counterpart, which aligns well with the analytical results available in the literature. The analytical expression of average age of information available in the literature were derived based on the assumption that no packets can be decoded when there is a collision. Our experiments demonstrated that the curves of the analytical expressions can match well with the corresponding excremental curves when the received powers of all IoT devices at the access point are tuned to be roughly the same. Nevertheless, when there exist imbalanced received powers, mismatches between analytical and experimental results appear, which is caused by the fact that the packet with higher received power can be decoded correctly even when it collides with a packet with lower received power. This phenomenon is referred to as ``capture effect''. This observation in our experiments motivates us to investigate the impact of capture effect on the average AoI of ALOHA-like random access protocols.

%\appendices
%\section{Proof of Lemma \ref{lemma2}}\label{app1}

\section*{Acknowledgment}
The authors thank Mr. Lihao Zhang for his constructive suggestions on the prototype development and thank Prof. Soung Liew for polishing the writing of the paper.
% cby themselves when using endfloat and the captionsoff option.
\ifCLASSOPTIONcaptionsoff
  \newpage
\fi

\bibliographystyle{IEEEtran}
\bibliography{References}

\end{document}

%% file: introduction.tex
\section{Introduction}
Internet of Things (IoT) represents one of the most significant paradigm shifts in recent decades poised to revolutionize several aspects of everyday life such as e-health and smart house. The new paradigm transforms every physical object into an intelligent entity capable of sensing, communicating and computing \cite{Lin2017IoTJ}. Ericsson foresaw that by 2021, there will be around 28 billion IoT devices and a majority of them will be empowered by wireless communication technologies \cite{Ericsson2016}. Analysts predicted that by 2025, the economic impact of the IoT could reach US \$11 trillion, or 11\% of the global economy, and by 2030 IoT could influence nearly the entire economy~\cite{McKinsey2015}.

%The applications of the IoT can be classified into three categories \cite{MIT2018how}: (1) consumer IoT, (2) business IoT, and (3) industrial IoT (IoT). As pointed out by the industrial giant GE, the IoT, also known as Industrial Internet, ``\emph{brings together brilliant machines, advanced analytics, and people at work. It is the network of a multitude of industrial devices connected by communications technologies that results in systems that can monitor, collect, exchange, analyze, and deliver valuable new insights like never before. These insights can then help drive smarter, faster business decisions for industrial companies}" \cite{GE_IoT}. IoT is transforming day-to-day operation in various domains, including manufacturing, smart cities, oil and gas, and utility. %Thanks to its great potential, IoT has attracted enormous attentions recently, see e.g., \cite{Mumtaz2017iem,Qiu2018tii,Hu2018Nonorthogonal,Sisinni2019tii,yue2019icc} and references therein.

%In this paper, we concern the design of wireless IoT networks, through which IoT devices transmit data to gateways and send the information to the cloud for remote processing.
A typical wireless IoT network is made up of three main ingredients: 1) IoT devices, 2) a communication network, and 3) gateways serving as information fusion nodes. The IoT devices are often deployed to observe the physical characteristic of a certain process, e.g., corrosion condition inside a refinery pipe, pollution levels within a gas factory, and speed and location of automated guided vehicles. The sensed data are transmitted through the communication network to the information fusion gateways where they are processed to extract useful information for the prediction, diagnostic, and decision making. For many applications, the usefulness of the information is heavily dependent on the freshness of the data measurements of the IoT devices when they arrive at the gateways~\cite{Shreedhar2018Acp}.

{Conventional performance metrics, e.g., throughput and delay, cannot adequately capture the information freshness.} Specifically, due to the random network delay, maximizing throughput or minimizing delay does not necessarily guarantee the freshest information observed at the receivers \cite{Sun2017tit}. In this context, the AoI concept was coined in \cite{Kaul2011mini} as a new metric to measure the information freshness at the destination side. AoI is a function of both how often packets are transmitted and how much delay that packets experience in the system. The metric of AoI is of great importance in time-sensitive IoT applications where the timeliness of information is crucial, and thus has attracted increasing attention recently, {see, e.g., \cite{kadota2018scheduling,wang2018skip,gu2019minimizing,maatouk2019minimizing,gu2019timely,wang2019minimizing,wang2020minimizing,wang2019minimizing2,li2020ageoriented,chen2020multiuser,chen2020secure} and references therein.}

With the new metric of AoI, a fundamental design problem for large-scale wireless IoT networks is ``\emph{how to schedule the status update packets of massive IoT devices in a decentralized manner towards achieving a low network-wide AoI}?". %\textcolor{red}{The reason why the decentralized scheme is more preferable is that it can not only save the resources used for systematical control but also maximize the network throughput or minimize the packet delay.}
%The design of such a scheduling policy is a non-trivial task. The reasons are three-fold: (1) Due to the broadcast nature of wireless signals, IoT devices cannot transmit status updates simultaneously. Otherwise, collisions occur and the status updates will not be delivered correctly. (2) Decentralized scheduling policy is preferred, because the network overhead in a centralized system can be significant, especially for massive IoT networks with a large number of connected devices. (3) To maintain low cost and low energy consumption for the IoT devices, the scheduling policy should be easily implemented and have low computation complexity.
%Though the analysis and optimization of AoI for various network setups have become an increasingly hot topic recently,
We note that there has only been limited work that attempted to answer this fundamental question \cite{S.Kaul-Distributed-Centralized-MAC,R.Talak-Distributed-MAC,Kosta2019age,chen2019age,chen2020aoi}. Specifically, \cite{S.Kaul-Distributed-Centralized-MAC,R.Talak-Distributed-MAC,Kosta2019age} investigated age-independent stationary randomized policies, in which each transmitter sends its packet with a certain fixed probability that can be optimized for minimizing the network-wide AoI ahead of time. %The work in \cite{jiang2018timely} designed a round-robin scheme for AoI minimization without considering the packet transmission errors.
%The follow-up work \cite{jiang2018decentralized} additionally assumed that nodes are provided with carrier sensing capabilities and proposed distributed schemes that have good performance in simulations; nevertheless, \cite{jiang2018decentralized} does not address how the parameters of the proposed algorithms should be designed.
Very recently, \cite{chen2019age} and \cite{chen2020aoi} devised age-dependent random access policies that can leverage the latest AoI at the transmitter and receiver sides to make age-based decisions at the transmitter side. The key difference between the random access policies proposed in \cite{chen2019age} and \cite{chen2020aoi} lies in how the channel access probability (CAP) of IoT devices is determined. Specifically, the CAP of each
device in \cite{chen2019age} is calculated based on conventional ALOHA backoff mechanisms, while the scheme in \cite{chen2020aoi} uses a pre-determined CAP for each IoT end device. Moreover, the random access policy proposed in \cite{chen2019age} is optimized only for the case when the number of devices approaches infinity. In contrast, the results presented in \cite{chen2020aoi} are suitable for any number of IoT devices. Nevertheless, all the aforementioned work focused on theoretical design and analysis. The development and implementation of a working prototype to evaluate and further improve these random access policies have been overlooked.

As the first attempt to fill such a gap, in this paper we build a prototype for testing and comparing the recently proposed AoI-oriented random access protocols. Specifically, we develop a testbed based on the Ettus USRP software-defined radio (SDR) equipment. To enable a time slotted system, we devise an over-the-air time synchronization scheme. We also design the frame structures of various packets exchanged within the system. Finally, we design the workflow of experiments, and test and compare the performance of the concerned protocols in an office environment.

Our experiment results show that the experimental average AoI curves of both age-independent and age-dependent schemes match well with their theoretical counterparts derived in the literature, when the signal powers at the AP's received antenna transmitted by different end devices are tuned to be roughly the same. The signal power at the AP's received antenna will be referred to as the received power hereafter. 
To unveil the impact of different levels of received powers of end devices at the AP, we adjust the received powers for all the IoT devices at the AP into two levels. That is, one group of IoT devices has higher received power than the other group. Experiment results show that the average AoI of the group with a higher received power has a lower average AoI, while the group with a lower received power has a higher average AoI. The rationale behind this observation is that the status update packets from the group with a stronger received power may still be decoded successfully by the AP even when they collide with the status update packets from the other group. Furthermore, as the difference between the received power for the two groups increases, the gap between their average AoI tends to be stable.

The rest of the paper is organized as follows. In Sec. II, we describe some preliminaries on the system model, the definition of AoI, and the principles of the considered AoI-oriented random access protocols. Sec. III elaborates on the development of the SDR-based prototype. In Sec. IV, we present and discuss the experimental results. Finally, Sec. V concludes the paper.

%% file: system_model.tex
\section{Preliminaries on System Model, AoI, and Protocols}
%\begin{figure}
%\centering \scalebox{0.45}{\includegraphics{multisystem.pdf}}
%\caption{The considered IoT network with one AP and $N$ IoT devices. }\label{fig:systemmodel}
%\end{figure}
In this section, we present some preliminaries on the system model, the definition of AoI, and the principles of the considered AoI-oriented random access protocols.

\subsection{System Model}
Consider an uplink IoT network consisting of an AP serving as the gateway and $N$ IoT devices denoted by $D_1,D_2,\cdots,D_N$. All the IoT devices aim to report their statuses as timely as possible to the AP via a common wireless channel. Time is divided into slots of equal durations and the transmission of each status update packet takes exactly one time slot. All IoT devices implement a slotted ALOHA-like random access protocol. Specifically, during each time slot, each IoT device can become either active or inactive according to a probability. If the device $D_i$ is active during one time slot, it firstly samples the latest state of the underlying process and generates a status update packet at the beginning of the time slot, which is known as the ``generate-at-will" model in the literature. The device $D_i$ then sends the generated status update packet to the AP. Otherwise, if the device $D_i$ chooses to be inactive, it stays idle during the time slot. Moreover, we assume that collisions happen if more than one device becomes active during the same time slot, and the AP cannot receive the status update packets correctly in collisions.

%In the following two subsections, we first formally define the average AoI, and then describe the proposed age-dependent random access (ADRA) protocol.
%The generated status updates of the IoT devices reflect the current status information of them, and the AP is responsible for keeping the latest condition of all the IoT devices as timely as possible.
\subsection{{Definition of the AoI}}\label{AoIdefinition}
%\begin{figure}
%\centering \scalebox{0.35}{\includegraphics{AoI.pdf}}
%\caption{One possible evolution of the instantaneous AoI for the device $D_i$ versus time slot index \cite{chen2020aoi}. }\label{fig:AoI}
%\vspace{-1em}
%\end{figure}
In this paper, the timeliness and freshness of the status update packets from the IoT devices at the AP are quantified by the recently proposed AoI metric. Denote by $t=1,2,3,\cdots$ the index of time slots and denote by $\Delta_i\left(t\right)$, $i\in\left\{1,2,\cdots,N\right\}$, the instantaneous AoI of the $i$-th IoT device in the time slot $t$. We then can mathematically express $\Delta_i\left(t\right)$ as \cite{Kaul2011mini}
\begin{equation}
\Delta_i\left(t\right)= t - \mu(t),\
\end{equation}
where $\mu(t)$ is the generation time of the most recently received status update packet in the time slot $t$.

We use $I_i\left(t\right)$ to denote the indicator of the active or inactive status for device $D_i$ in the time slot $t$. Particularly, $I_i\left(t\right)=1$ means that the device $D_i$ is active during the time slot $t$. Otherwise, $I_i\left(t\right)=0$. Based on the definition of the AoI, the instantaneous AoI of the device $D_i$ drops to one when the device $D_i$ is active and all the other devices are inactive, i.e., the device $D_i$ successfully delivers a status update packet to the AP. Otherwise, the instantaneous AoI of the device $D_i$ increases by one for each time slot. Mathematically, the evolution of the instantaneous AoI for the device $D_i$ can be expressed as
\begin{equation}
\Delta _i \left( {t  + 1} \right) =\left\{
\begin{matrix}
\begin{split}
   &{1}, \quad\text{if}\quad I_i \left( t  \right)=1, I_j \left( t  \right) = 0, \forall j\ne i\\
   &{\Delta _i \left( {t } \right)+1}, \quad\text{otherwise}  \\
\end{split}
\end{matrix}
\right..
\end{equation}
%To ease understanding, the readers are referred to \cite[Fig. 1]{chen2020aoi} that illustrates the evolution of the instantaneous AoI of $D_i$ for 14 consecutive time slots with a starting value of 1.
{Based on the AoI evolution, the long-term average AoI for the $i$-th IoT device can be defined as}
\begin{equation}\label{AoIexpression}
\bar \Delta _{i}  = \mathop {\lim }\limits_{T \to \infty } {1 \over T}\sum\limits_{t  = 1}^T {\Delta _i \left( t  \right)}.
\end{equation}
\subsection{Principles of AoI-oriented Random Access Protocols}
We now introduce the principles of the ALOHA-like AoI-oriented random access protocol proposed in \cite{chen2020aoi}.
\subsubsection{\textbf{Age-Independent Random Access}}
In the AIRA protocol, devices access the channel with the same probability no matter whether their instantaneous AoI values are low or high. The AIRA protocol is easy to implement in a distributed manner. However, it has the shortcoming of not leveraging the instantaneous AoI information at the transmitter side \cite{chen2019age}. The performance of the AIRA protocol was analyzed and optimized in \cite{S.Kaul-Distributed-Centralized-MAC}. More specifically, the average AoI of the considered $N$-device network can be expressed as~\cite{S.Kaul-Distributed-Centralized-MAC}
{\begin{equation}\label{eq:AIRA}
\bar \Delta_{\rm AIRA}  = {1 \over {p \left( {1 - p } \right)^{N - 1} }},
\end{equation}
where $p$ is the fixed CAP of all IoT devices. The optimal CAP that minimizes the network wide average AoI is given by $p^*_{\rm AIRA} = 1/N$.
\begin{figure}
\centering \scalebox{0.35}{\includegraphics{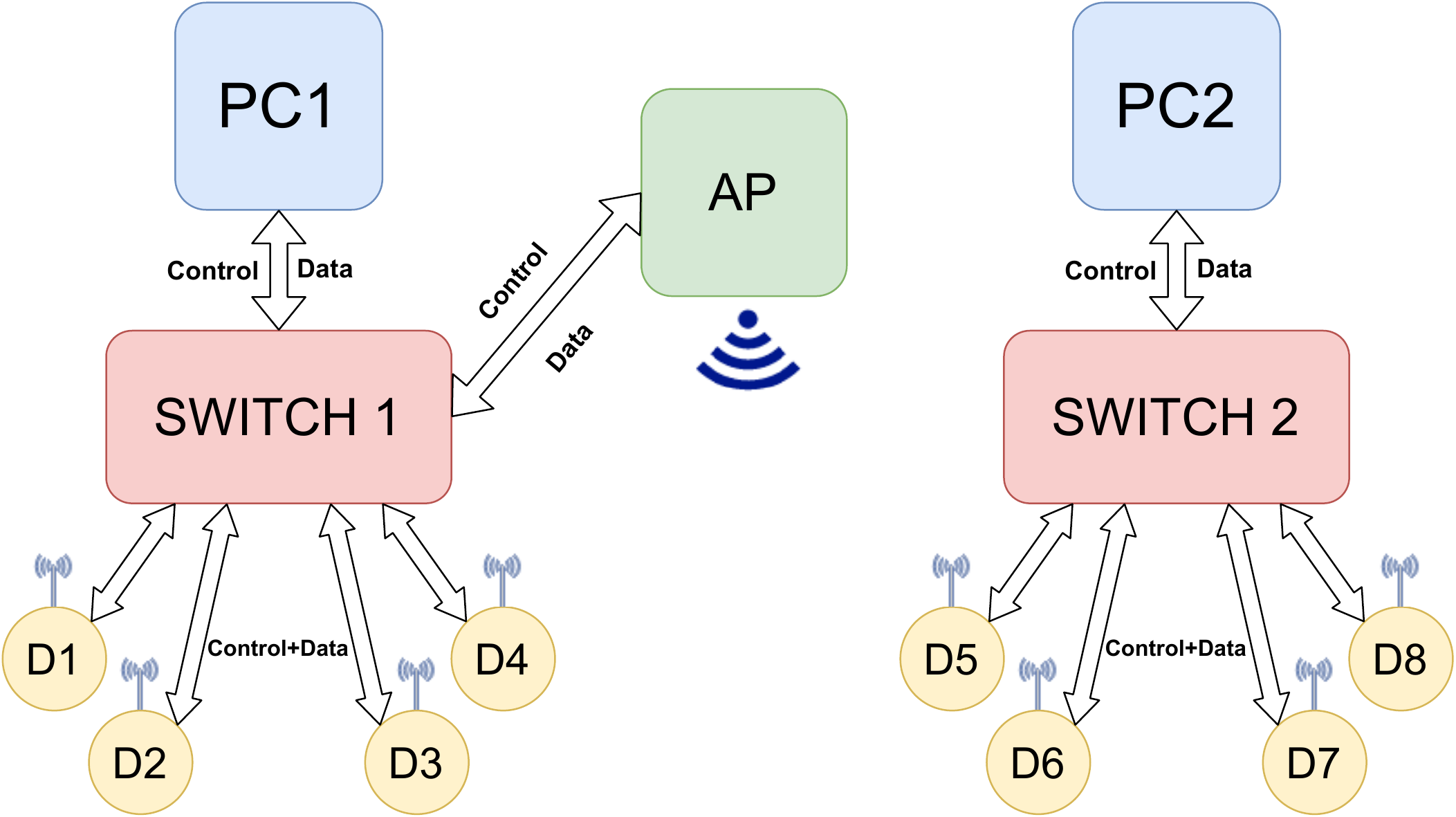}}
\caption{A block diagram of the connections among the USRPs and PCs in the developed prototype, in which the double-headed arrow $\Longleftrightarrow$ represents 1G Ethernet cables.}\label{fig:connection}
\vspace{-1em}
\end{figure}
\subsubsection{\textbf{Age-Dependent Random Access}}
To improve the system performance by properly leveraging the instantaneous AoI information available at the transmitter side, \cite{chen2020aoi} developed an ADRA protocol. In the ADRA protocol, the CAP of each device is no longer a constant. Instead, the CAP varies according to the instantaneous age of the IoT devices. A key design problem for the ADRA protocol lies in ``\emph{how to adjust the CAP as the instantaneous AoI values evolves}?''. A simple yet effective strategy is to impose an age threshold such that those devices with instantaneous age being not larger than the threshold will keep silent, and those with instantaneous age exceeding the threshold will access the channel with a certain probability. In this paper, we focus on the implementation and evaluation of the ADRA protocol with a fixed threshold that was referred to as a threshold-based ADRA protocol in \cite{chen2020aoi}, which has been shown to have a better performance than the one proposed in \cite{chen2019age}.

In the ADRA protocol from \cite{chen2020aoi}, all IoT devices maintain a fixed age-dependent CAP vector $\mathbf{p}=\left\{p_1,p_2,p_3,\cdots, \right\}$, where $p_l$ denotes the active probability when the instantaneous AoI is equal to $l$. Specifically, if the instantaneous AoI is no less than the threshold $\delta$, the IoT device transmits with a fixed probability $p$. Otherwise, the IoT device keeps idle with probability 1. Thanks to the simplified CAP model, an approximate closed-form expression of the average AoI for the ADRA protocol can be obtained and be given as \cite{chen2020aoi}
{\begin{equation}\label{eq:ADRA}
\bar{\Delta}\approx{\frac{\delta}{2}+\frac{1}{p q}-\frac{\delta}{2(\delta p q+1-p q)}}.
\end{equation}

%% file: expriment.tex
\section{A Software-Defined Radio-based Prototype}
In this section, we elaborate on the development of a proof-of-concept prototype by using a stream-oriented real-time SDR platform called GNURadio to test and verify the performance of the considered random access protocols. We use the Ettus USRP N210's as transceivers: One USRP N210 serves as the AP and eight USRP N210s serve as the IoT devices.
\begin{figure}
\centering \scalebox{0.04}{\includegraphics{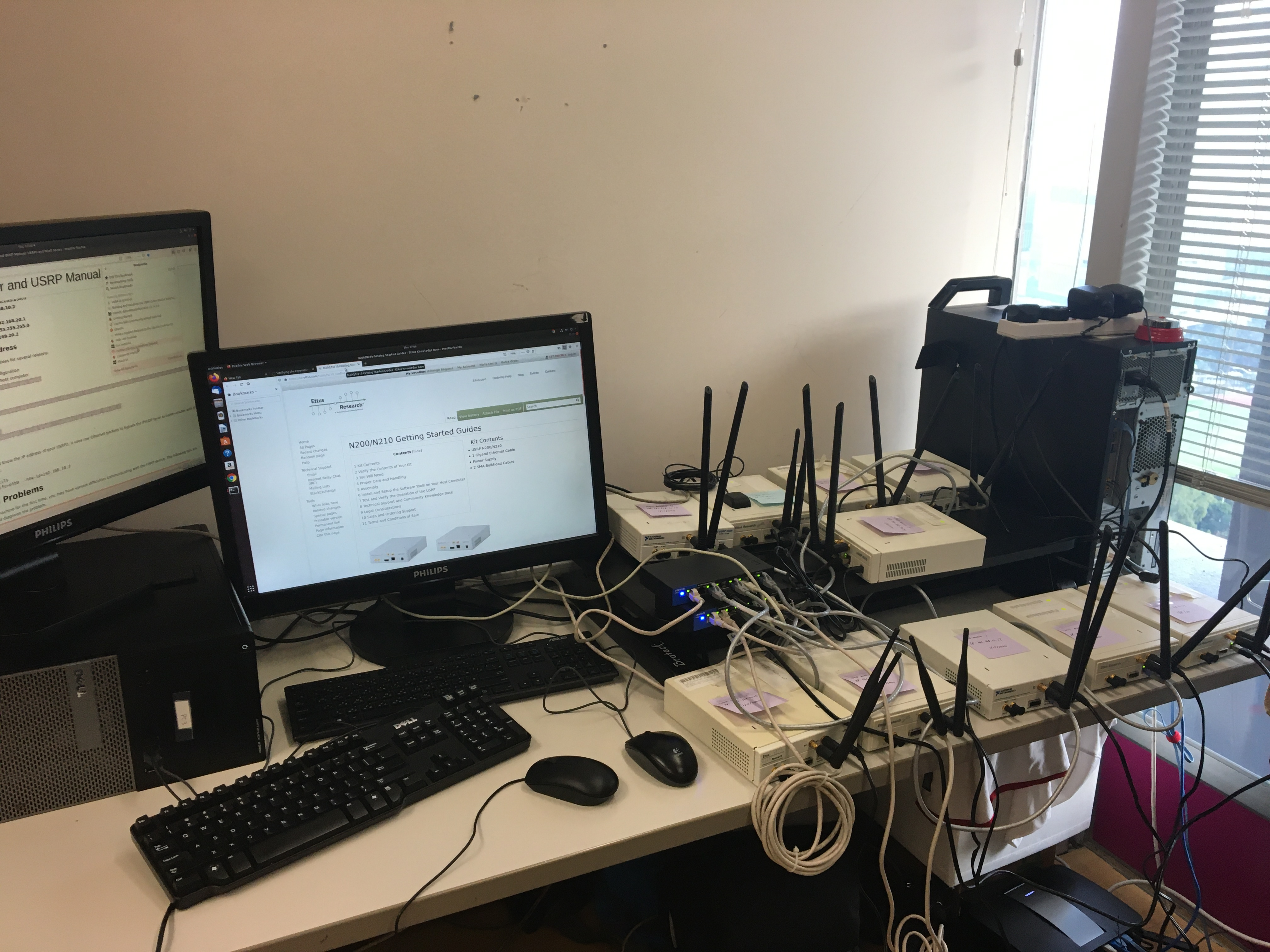}}
\caption{A photograph of our prototype.}\label{fig:platform}
\vspace{-1em}
\end{figure}All of them are connected to two powerful PCs\footnote{In principle, each USRP equipment can be connected to a dedicated PC for baseband signal processing. We connect multiple USRPs to one PC for saving the prototype cost.} through multiple 1 Gigabit Ethernet cables and two Ethernet switches. A block diagram of the connections among the USRPs and PCs is given in Fig. \ref{fig:connection}. It is worth emphasizing that though some USRPs are connected to the same PC, they use individual local clocks on their motherboards, and thus the time synchronization among all devices is needed to guarantee that every device accesses the channel medium in a slotted manner. The SBX RF front-boards \cite{daughter} embedded in USRPs are used to transmit RF signals, working at 1 GHz with 500 kHz channel bandwidth. We use a program developed in the GNURadio platform to define the signal generation and data processing in our SDR prototype. GNURadio is a mature software framework that provides a comprehensive library of signal processing blocks. With this powerful tool, users are able to design and deploy a real-world radio system on USRPs according to their own demands \cite{gnuradio}. A photograph of the developed proof-of-concept prototype is shown in Fig. \ref{fig:platform}. To ensure high transmission reliability when there is no collision as assumed in the previous section, we implement the convolutional code in the physical layer. Orthogonal frequency duplex modulation (OFDM) is used for higher frequency efficiency.

\begin{figure*}
\centering \scalebox{0.4}{\includegraphics{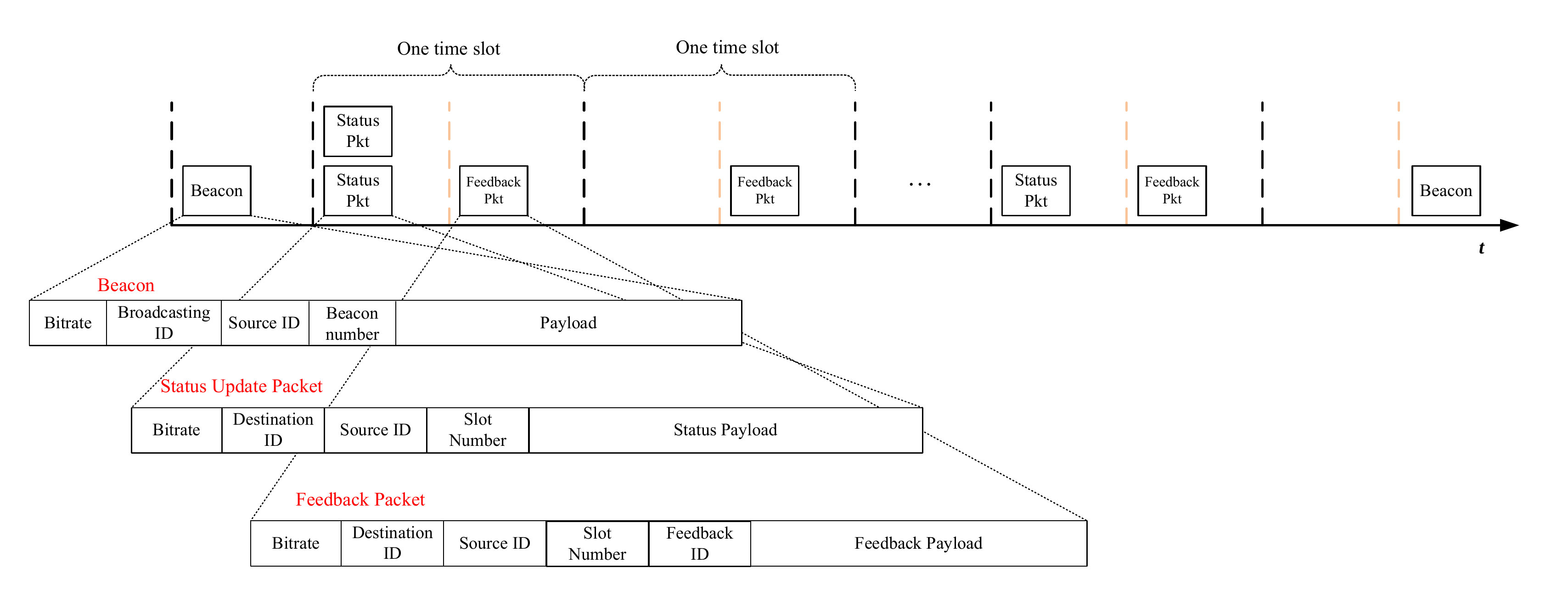}}
\caption{A diagram of the transmission scheme as well as the frame structures of {three} types of packets in our experiments, where all IoT devices remain inactive in the second time slot.}\label{fig:frame}
\vspace{-1em}
\end{figure*}
To realize time-slotted transmissions among the USRPs, we design a simple yet effective synchronized transmission scheme. The scheme involves three types of transmissions: (1) beacon broadcasting, (2) channel access, {and (3) feedback}. The beacon broadcasting phase aims to achieve the time synchronization. In this phase,
the USRP that acts as the AP broadcasts a Beacon, which contains timing information to serve as a time reference for a slotted communication scenario. The duration of the inter-beacon period is determined by the precision of the USRP oscillator, which is set to 100 time slots in our experiments. In our prototype, each time slot is further split into two transmission slots, which are used to convey Status Update Packets and Feedback Packets, respectively. The channel access phase starts at the beginning of the subsequent transmission slot. The USRP(s) operating as IoT device(s) sends the Status Update Packet(s) to the AP if it chooses to be active according to the age threshold and CAP. Immediately following the IoT device(s)' transmission slot, the AP broadcasts a Feedback Packet indicating the successful reception of a Status Update Packet. An IoT device only resets its AoI back to one if its identity is declared in the received Feedback Packet. Otherwise, its AoI increases by one.

For a better understanding, a diagram of the transmission scheme as well as the frame structures of three types of packets are depicted in Fig. \ref{fig:frame}. The functions and structures of these frames are elaborated in the following:
	
\begin{itemize}
  \item \textbf{Beacon} is sent by the AP and carries the time synchronization information. It includes the following parts
      \begin{itemize}
        \item \emph{Bitrate} gives the modulation bitrate information.
        \item \emph{Destination ID} is the broadcasting ID.
        \item \emph{Source ID} gives the AP's node ID.
        \item \emph{Beacon number} is the current beacon's number.
        \item \emph{Payload} carries the interval information.
      \end{itemize}
  \item \textbf{Status Update Packet} is sent by IoT devices. It is used for reporting the devices' statuses or the related information to the AP.
      \begin{itemize}
        \item \emph{Bitrate} gives the modulation bitrate information.
        \item \emph{Destination ID} is the AP's node ID.
        \item \emph{Source ID} gives the transmitting device's node ID.
        %\item \emph{Recv Beacon number} is the counting number of the beacon which triggers the current Status Update Packet.
        {\item \emph{Slot number} is the counting number of the current time slot.}
        \item \emph{Status Payload} provides the device's status information.
      \end{itemize}

  \item \textbf{Feedback Packet} is broadcasted by the AP. It is used for notifying IoT devices about the reception of Status Update Packet in the previous transmission slot, which allows each IoT device to calculate its instantaneous AoI.

  	 \begin{itemize}
  	 	\item \emph{Bitrate} gives the modulation bitrate information.
  	 	\item \emph{Destination ID} is the broadcasting ID.
  	 	\item \emph{Source ID} gives the AP's node ID.
  	 	%\item \emph{Recv Beacon number} is the counting number of the beacon which triggers the current Status Update Packet.
  	 	\item \emph{Slot number} is the counting number of the current time slot.
  	 	\item \emph{Feedback ID} is the Node ID who successfully transmitted a Status Update Packet in previous transmission slot.
  	 	\item \emph{Feedback Payload} provides other control information for the IoT devices.
\end{itemize}

\end{itemize}

\section{Experiment Results and Discussions}

In this section, we will present the experimental average AoI of the ADRA and AIRA protocols obtained by using our prototype. We then discuss the impact of imbalanced received powers (i.e., the received powers of IoT devices are set to different levels) on the AoI performance.
 \begin{figure}
    \centering
    \small

    \includegraphics[width=0.42\textwidth]{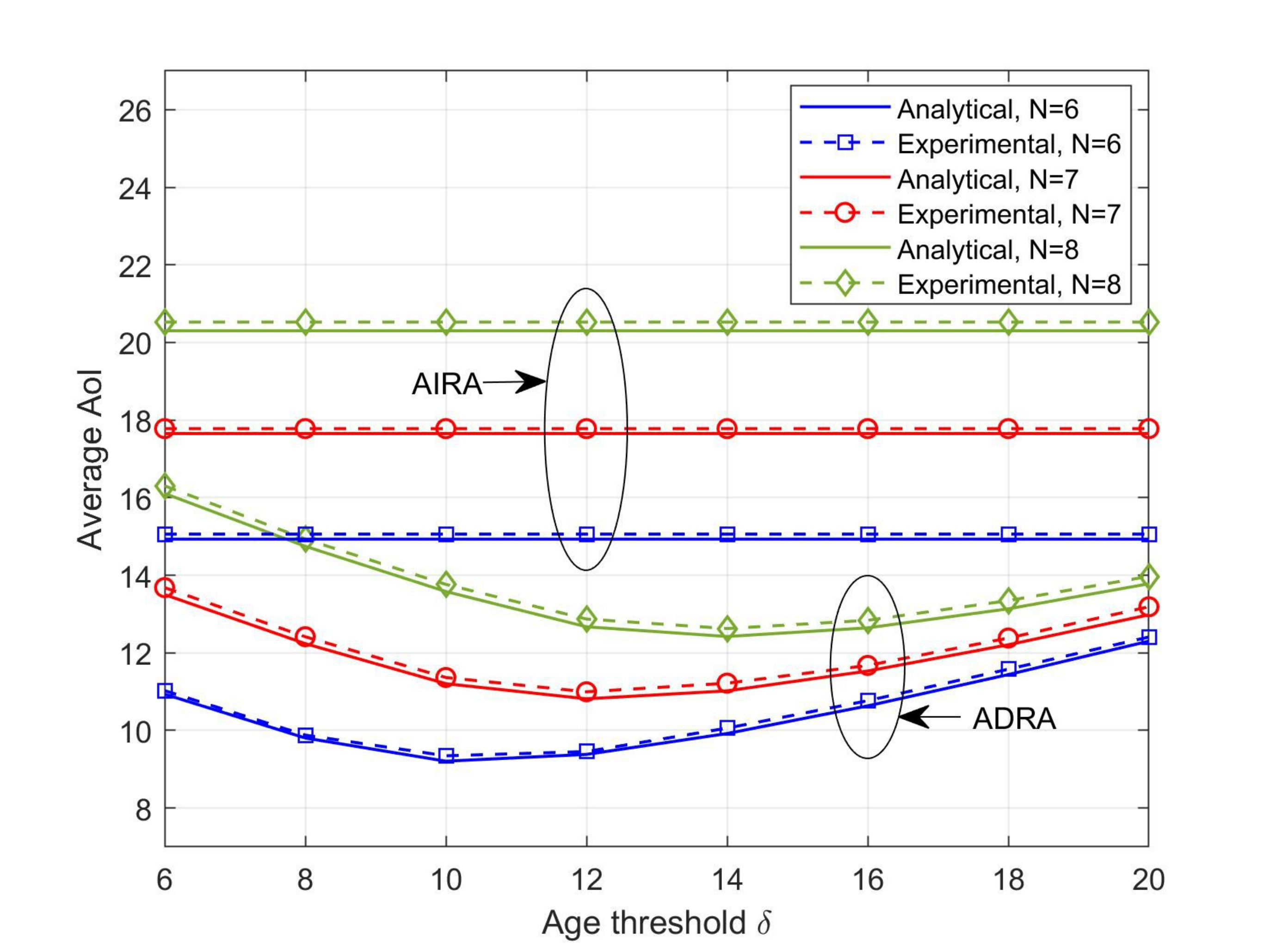}

    \caption{Comparison of the experimental and analytical results of the ADRA and AIRA protocols when the number of IoT devices are equal to 6, 7, 8.}
   \label{fig:experiment_result_1}
\end{figure}
\subsection{Performance in A Small-Scale Network Setup}
We first evaluate and compare the performance of AIRA and ADRA protocols in an IoT network with a small number of devices. To that end, we conduct three experiments by varying the number of USRPs serving as the IoT devices from 6 to 8. For each experiment setup, we first adjust the transmit powers of the USRPs representing the IoT devices in the GNURadio program to ensure that their received powers at the AP are nearly the same. We then gradually change the values of the age threshold. For each age threshold, we find the optimal CAP value by utilizing the analytical results of the ADRA protocol derived in Sec. III of \cite{chen2020aoi}. Since the CAP in the AIRA protocol is not relevant to the instantaneous AoI, its age threshold and CAP values are fixed as 1 and $1/N$, respectively. We next use these parameters to conduct packet transmissions while the host PCs connected to each IoT device record the AoI evolution. The experiment takes a duration of 800 time slots, and at the end of the experiment, we average the AoI for all the IoT devices according to the records in the host PCs. We also calculate the analytical average AoI of the ADRA and AIRA protocols using the mathematical expressions given in (\ref{eq:AIRA}) and (\ref{eq:ADRA}).

The experimental and the corresponding analytical results are shown in Fig. \ref{fig:experiment_result_1}. We can see that for both protocols, the analytical and experimental results in terms of the average AoI are close to each other in each experimental case with the difference between them being no more than 0.3. The slight mismatches between them are caused by the packet misdetection when no collision happens. Note that we conduct the experiment in an office environment and our experiments show that the packet misdetection rate is around 2\%, which could be caused by the USRP hardware imperfection.
\begin{figure}
    \centering
    \small

    \includegraphics[width=0.42\textwidth]{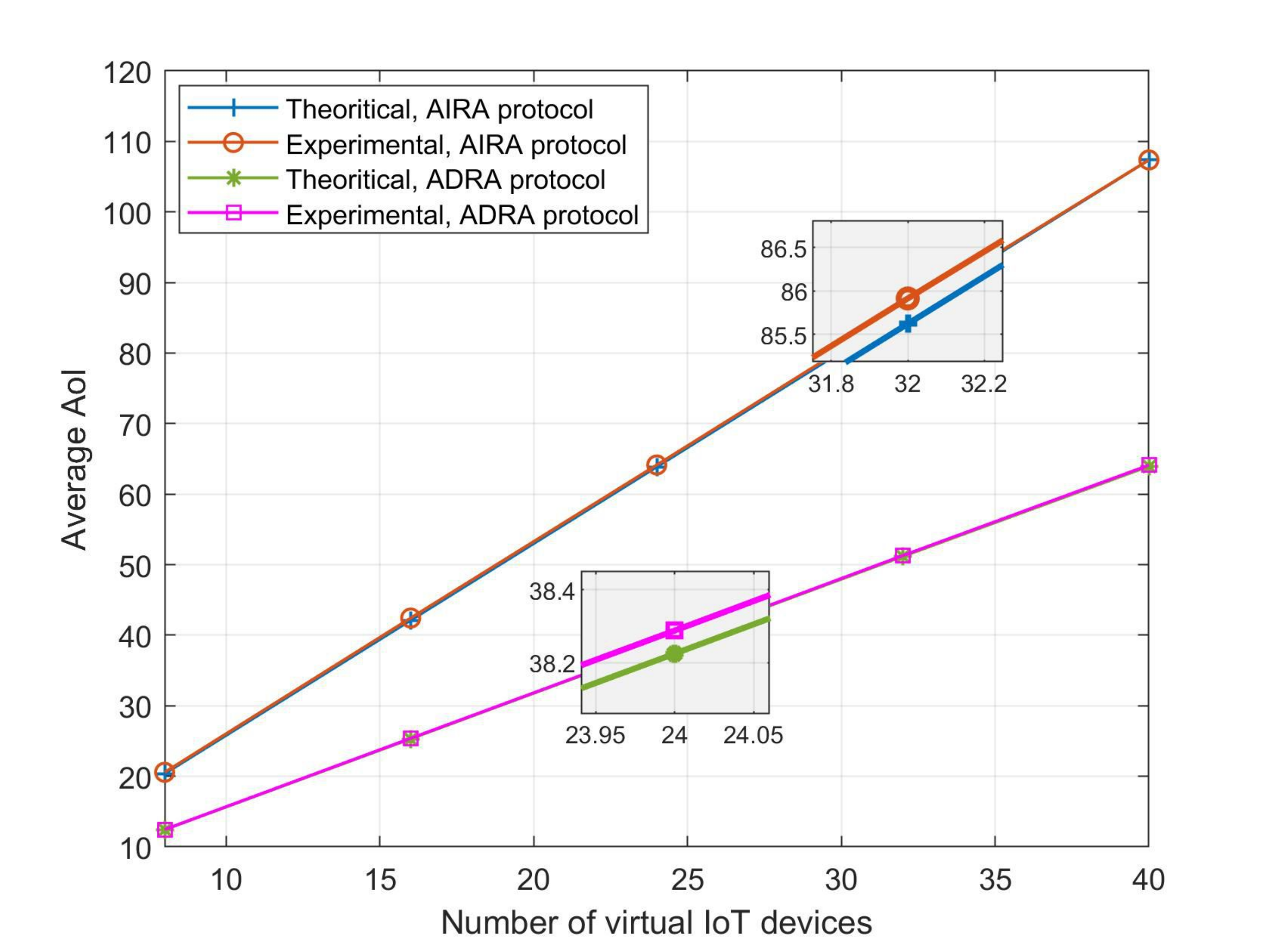}

    \caption{The experimental and analytical average AoI versus the number of IoT devices for the ADRA and AIRA protocols when the number of IoT devices ranges from 8 to 40.}
   \label{fig:experiment_result_2}
\end{figure}

\subsection{Performance in A Large-Scale Network Setup}
To overcome the limitation on the number of available USRPs in our laboratory, we implement a new mechanism in the GNURadio program to drive one USRP for acting as multiple virtual IoT devices. The idea is to use one USRP to represent multiple IoT devices and emulate a part of transmission before ejecting Status Update Packets into the channel. At the beginning of each time slot, each virtual IoT device checks whether its AoI value exceeds the predefined age threshold. If not, the virtual IoT device keeps silent. Otherwise, it becomes active and tries to access the channel according to the CAP. For an USRP driven as multiple virtual IoT devices, if there is only one virtual IoT device chooses to be active, it will transmit a Status Update Packet to the AP. If not, it remains silent as an internal collision occurs. After the USRP receives a Feedback Packet from the AP, it will reset the AoI of the virtual IoT device whose identity is declared in the Feedback Packet back to one while increasing the AoI of the other devices by one. If there is no broadcasted Feedback Packet in the channel, the USRP will increase the AoI of all its underlying virtual IoT devices by one.
\begin{figure}
    \centering
    \small

    \includegraphics[width=0.42\textwidth]{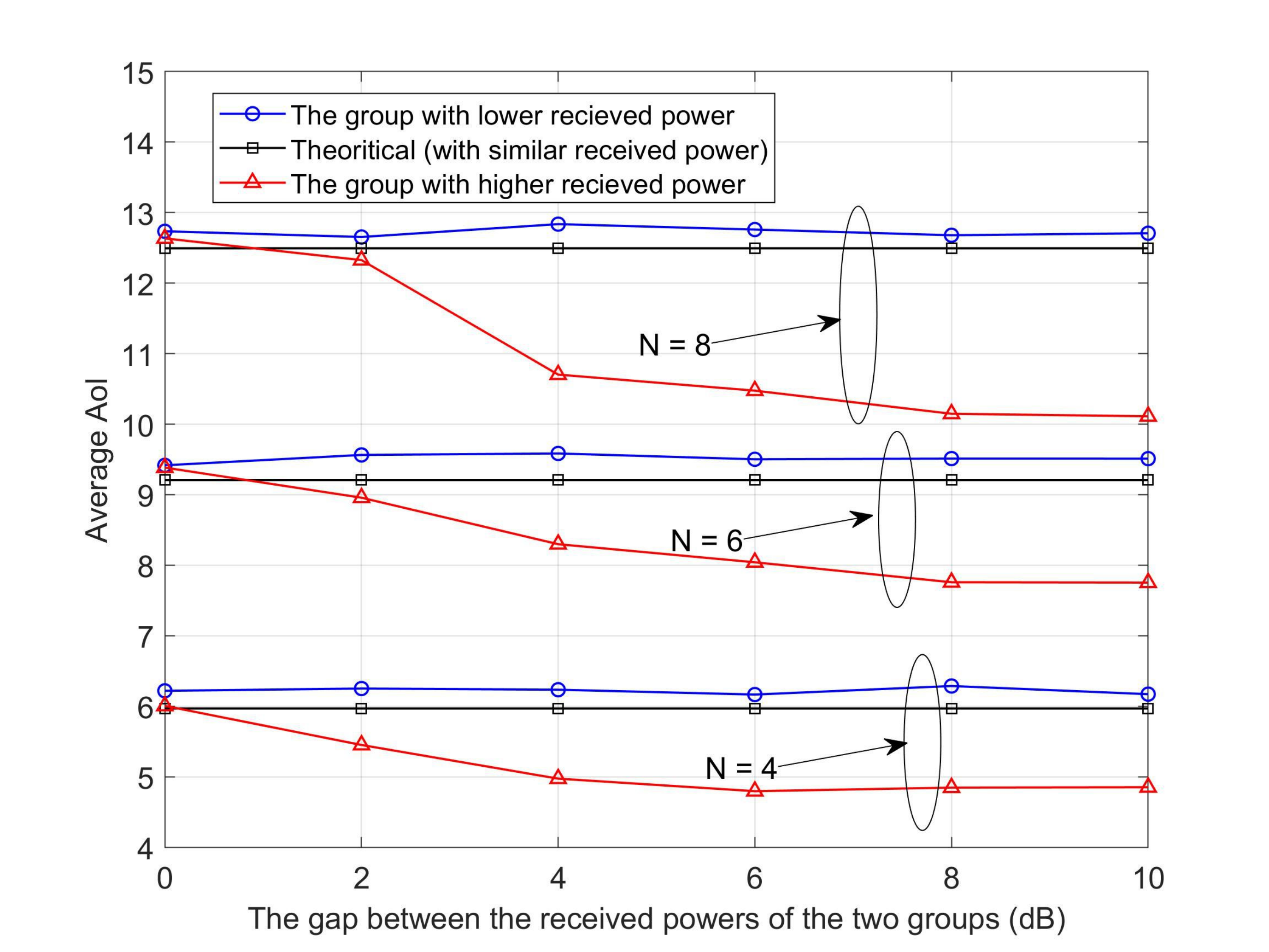}

    \caption{Comparison of the experimental results between IoT devices with different received signal strength at the AP.}
   \label{fig:experiment_result_3}
\end{figure}

With the above improved GNURadio program, our prototype becomes capable to mimic larger IoT networks with more than 8 IoT devices. We then design an experiment to compare the performances of the AIRA and ADRA protocols in network setups with 8 to 40 virtual IoT devices. In each new experiment, the number of the virtual IoT devices increases by 8. We then conduct packets transmissions over more than 800 time slots and obtain the average AoI of the AIRA and ADRA protocols, respectively. At the end of the experiment, we again attain the analytical average AoI for both the AIRA and ADRA protocols by using the mathematical expressions given in (\ref{eq:AIRA}) and (\ref{eq:ADRA}).

The experimental and analytical results are plotted in Fig. \ref{fig:experiment_result_2}. We can see that the experimental average AoI of the AIRA and ADRA protocol is still close to the analytical value in a larger IoT network with more than 10 nodes. Comparing Fig. \ref{fig:experiment_result_2} with Fig. \ref{fig:experiment_result_1}, we can find that the average AoI of the ADRA protocol is always smaller than that of the AIRA protocol for the simulated cases. Besides, the gap between the average AoI of these two protocols increases with more virtual IoT devices deployed in the network. The reason for this observation is that the ADRA protocol provides more opportunities to IoT devices with higher AoI values to access the channel, and thus the IoT devices co-exist in a more harmonious way, leading to lower network-wide average AoI performance.

\subsection{Impact of Imbalanced Received Powers}
In order to quantify the impact of different levels of received powers of IoT devices at the AP, we design three experiments with 4, 6 and 8 USRPs serving as IoT devices, respectively. In each experiment setup, we initially adjust the received powers of all IoT devices to be around 35dB at the AP. We then divide IoT devices into two groups and intentionally set their received powers into different levels. Specifically, we fix the received power of the lower group and then gradually increase the value of the received power for the higher group. 

Fig. \ref{fig:experiment_result_3} shows the experimental and analytical average AoI results. Fig. \ref{fig:experiment_result_3} shows that when the received powers of IoT devices are equal, the two groups have approximately the same experimental average AoI. which are close to the analytical counterpart. AS the gap between the received powers of the two groups enlarges, Status Update Packets from the group with a stronger received power have a higher probability to be detected by the AP even under the collision with packets from the other group. As a result, the average AoI of the group with higher received power gradually decreases. After the gap between the received powers is larger than 8dB, the average AoI of the group with a higher received power tends to remain stable. This is because that the AP can easily decode the Status Update Packets correctly from the group with a higher received power even under the interference with the other lower group. Keep increasing the gap thus will not benefit the group with a higher received power further. For the group with a lower received power, because the AP can hardly decode its Status Update Packets correctly under the interference from the group with a higher received power, its average AoI is almost the same for different gaps.